\documentclass{appolb}

\usepackage{lineno}
\usepackage{subfigure}
\usepackage{adjustbox}
\usepackage{amsmath}  
\usepackage{amsfonts} 
\usepackage{graphicx} 
\usepackage{multirow}
\usepackage{gensymb}
\usepackage{lipsum}
\usepackage{textcomp}
\usepackage[table]{xcolor}
\usepackage{soul}
\usepackage{cancel}
\usepackage{authblk}

\RequirePackage[colorlinks,citecolor=blue,urlcolor=blue,linkcolor=blue]{hyperref}
\RequirePackage{url}

\begin{document}
\title{Study of a large array to detect ultra high energy tau-neutrino }
\author[1]{S. ATIK YILMAZ 
\thanks{\texttt{seyma.atikyilmaz@gmail.com}}
}
\author[2]{A.YILMAZ 
\thanks{\texttt{aliyilmaz@giresun.edu.tr}} 
}
\author[3]{M. IORI  
\thanks{\texttt{maurizio.iori@roma1.infn.it}}
} 
\author[1]{ K. Y. OYULMAZ 
 \thanks{\texttt{kaan.oyulmaz@gmail.com}}
  }
  \author[1]{H. DENIZLI  
\thanks{\texttt{denizli\_h@ibu.edu.tr}}
}

\affil[1]{Bolu Abant Izzet Baysal University, Bolu 14280, TR}
\affil[2]{Giresun University, Giresun 28200, TR} 
 \affil[3]{La Sapienza, University of Rome, Rome 00185, IT} 

\maketitle
\begin{abstract}

The $PeV$ cosmogenic neutrino is still interesting argument. Since cosmogenic neutrinos interact weakly with matter, the detection of their direction will precisely point out the source in the space. 
In this paper, we show the results of the simulation of tau lepton air showers induced by high energy neutrinos detected by an array of stations designed to use the Earth Skimming method improved by the ``mountain chain screen" strategy.
Both track  time stamp and position information of the stations on the array are used to reconstruct the
shower to estimate the direction and the number of events.
The array studied consists of 640 stations ($40 \times 16$) spread over an area of $0.6\,km^2$ starting 
from $1500\,m$ above the sea level (a.s.l.) on $30\degree$ inclined plane of the mountain.
When we extrapolate to    3 years and 10  $km^2$  we estimate 13 tau lepton events in energy interval of 10 PeV to 1000 PeV detected using the present upper 
limits of tau neutrino flux.

\end{abstract}


\newpage
\section{Introduction}
\label{sec:intro}

The observation of Ultra High Energy (UHE) neutrinos generated in both distant galactic and extra galactic  sources such as pulsars
or supernova remnants have been a major challenge in astroparticle physics. 
The neutrinos with an energy of $10^{17}$ eV or more, generated by interaction of
Ultra High Energy Cosmic Rays (UHECRs) with the Cosmic Microwave Background (CMB)
are called cosmogenic neutrinos \cite{HALZEN1992184}.
Because of the low neutrino flux and the small detection probability, a very large scale detector
array (surface or volume) is required to get a detectable rate.
For this purpose, many large array experiments based on different techniques have been constructed 
\cite{Linsley1962, KAKIMOTO, BARWICK201512, Ackermann_2008, SOKOLSKY201174, IceCube2012, 1202.1493v1}.

Recently the IceCube Collaboration \cite{Aartsen2013} has reported 28 
neutrino candidate events in the energy range of $30$ to 1200 TeV. 
They also claimed the detection of a high-energy neutrino event,
IceCube-170922A, with an energy of $\sim290$ TeV  in correlation with other telescopes showing a multimessenger physics.
The arrival direction of this neutrino was consistent 
with the location of a known $\gamma$-ray blazar, TXS 0506+056, observed to be in a flaring state \cite{Multi}.

The cosmogenic neutrino, such as tau neutrino ($\nu_{\tau}$), may interact along the chord of the Earth or 
inside a mountain chain and creates a tau lepton ($\tau$) by the charged-current
(CC) interaction. When the $\tau$ emerges from the rock before it decays, it produces an extensive air shower developing at very large angle near to the horizon~\cite{Zas,Fargion99}. The observations can be achieved using surface detectors, Cherenkov light telescopes,
fluorescence detectors or radio signals. Some of the proposed experiments in the literature are Ashra-1~\cite{ASAOKA20137}, MAGIC telescopes~\cite{Gora:2017pre}, Telescope Array (TA)~\cite{cao2005ultra}, and  IceCube~\cite{ICRC2019_nu_ice3, ICRC2019_ICE_cube_res}. 
 
 The Ashra-1 detector aims to observe the Cherenkov lights
from $\tau$ showers with Earth Skimming method for PeV-EeV energy range \cite{ASAOKA20137}.
An upper limit of neutrino flux has been measured by MAGIC telescopes in the 1 PeV to 3 EeV energy interval for $\nu_{\tau}$ induced showers arising from the ocean \cite{Gora:2017pre}.
In Antartica, radio antennas are placed to the ground and point at the high mountains
to detect the skimming $\nu_{\tau}$s too~\cite{Taiwanese}. All these proposed experiments have no evidence of $\tau$ events except a possible $\tau$ candidate in IceCube~\cite{ICRC2019_nu_ice3, ICRC2019_ICE_cube_res}.
To  detect the horizontal and upward $\nu_{\tau}$ showers, we propose a surface detector array on an inclined plane to improve the detection acceptance of $\tau$ shower using the time of flight (TOF) and e/mu separation ~\cite{IORI2012285, Iori_2015}. In this study we also include a chain of mountain in front of the array and evaluate the wideness of the valley.

This paper shows the results of a large array named as ``TAUshoWER", \sloppy (TAUWER) which has the geometrical advantage of using both the Earth Skimming method to detect 
the $\tau$ showers  produced below the horizon and a mountain chain in front of the array to enlarge the acceptance. 
The array geometry is discussed in Section~\ref{sec:setup} while the simulation of  the decaying $\tau$ air showers,
the selection criteria for decay length from emerging point and the result of the probability calculations 
for tau-neutrino interaction are  given in Section~\ref{sec:simulation}. 
Analytical approach on the estimation of the event number to be observed with this array is presented in Section~\ref{sec:EventNumber}. Reconstruction of the showers is given 
in Section~\ref{sec:ArrDirection} and Section~\ref{sec:EnergyReconstruct}.
Section~\ref{sec:TriggEff} contains a discussion of the trigger. 
Finally, we summarize our results and conclude with number of expected event/year.

\section{Design of the Array}
\label{sec:setup}

\begin{figure}[!htb] 
 	\centering 
    		\includegraphics[width=1\textwidth]{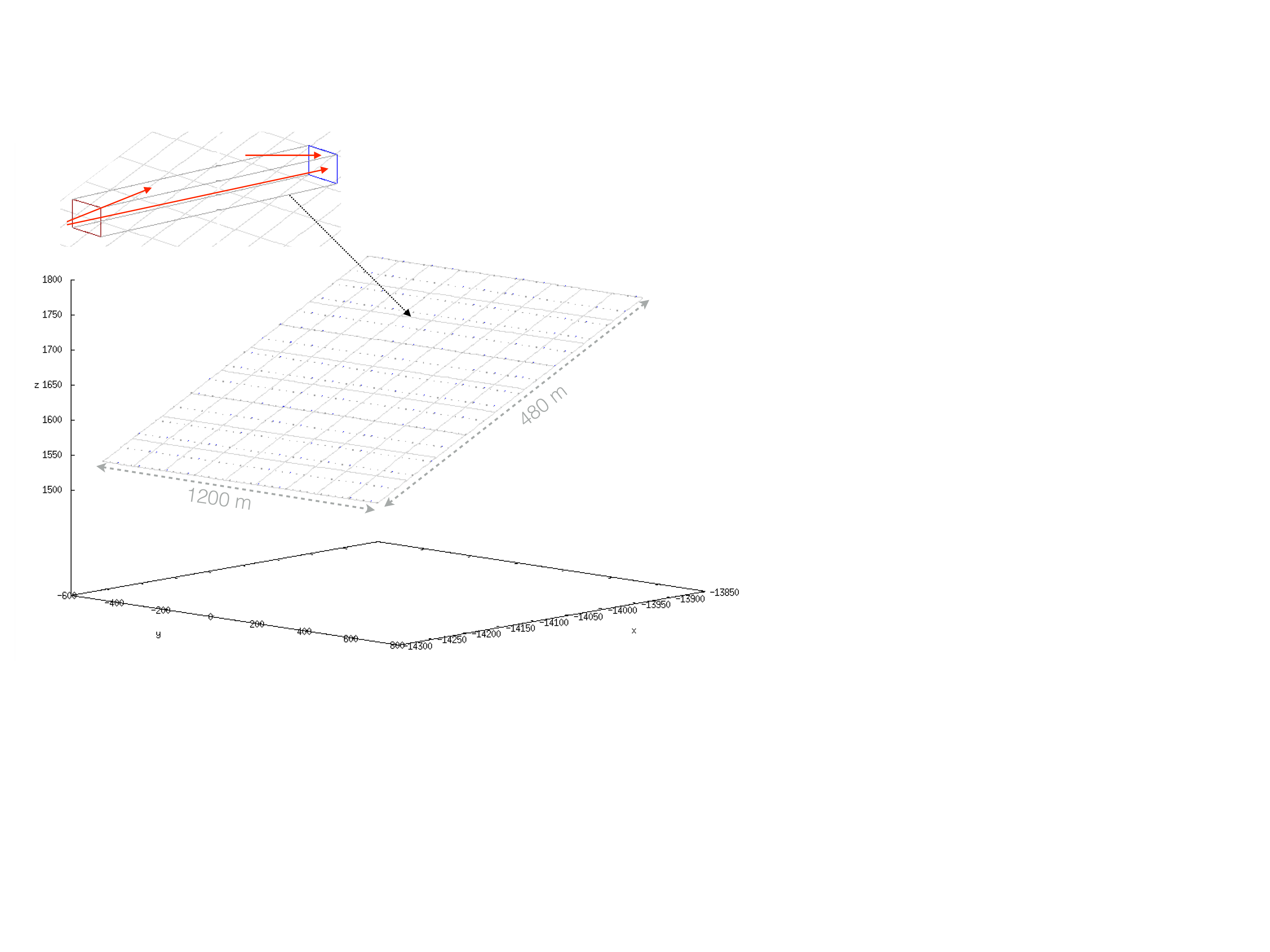}
    
  	\caption{Schematic drawing (scaled) of the array. The array consists of 640 stations 
  	placed on the $30\degree$ inclined plane (mountain surface) in the matrix form of 16 rows
  	 and 40 columns  which are separated by 30 m apart.}
	\label{fig:array}
\end{figure}

To detect the $\tau$ showers produced by the tau-neutrino interaction with rock along its path in the Earth crust or a mountain in front of the array, we propose a large  surface array in which stations point below the horizon to detect particles 
produced by large angle showers. The best way, as discussed in Section~\ref{sec:simulation}, to detect $\nu_{\tau}$ is to use large amount of  matter  by pointing down the horizon and a mountain chain in front of the array 
to increase the interaction probability of $\nu_{\tau}$.
Therefore, we propose to install stations on an inclined mountain surface.
In this paper, we consider a grid of 640 stations ($40 \times 16$) placed $30 \, m$ apart and located on $30\degree$ inclined plane of the mountain between $1500\,m$ and $2250\,m$ above the sea level and covering a surface of 0.6 km$^{2}$ as shown in Figure~\ref{fig:array}.
Each station, \textit{named tower}, is composed by two scintillator tiles ($40\times 20\times 1.5\, cm^{3}$) 160 cm apart and read by 3x3 mm$^{2}$ silicon photomultiplier (SiPM) placed on one side , which permit  to select the particle direction using the  time of flight (TOF) method.  If  a layer of lead 2.5 cm thickness is installed in front of the rear tile it is possible separate the muons from the electromagnetic particles~\cite{Iori_2015}.
The stations point below the horizon (approximately $2.5\degree$) to measure upward moving showers coming from $\nu_{\tau}$ interactions with the Earth crust and at the some time can detect showers produced in the mountain chain. Shower direction is selected by the TOF computed with the two scintillating plates. The layout of the array studied in this paper covers a surface of  0.6 km$^2$  which can be increased according to the shape of the mountain.

\section{Evaluation of $\tau$ events escaped from the rock}
\label{sec:simulation}

\begin{figure}[h] 
 	\centering 
    		\includegraphics[width=1\textwidth]{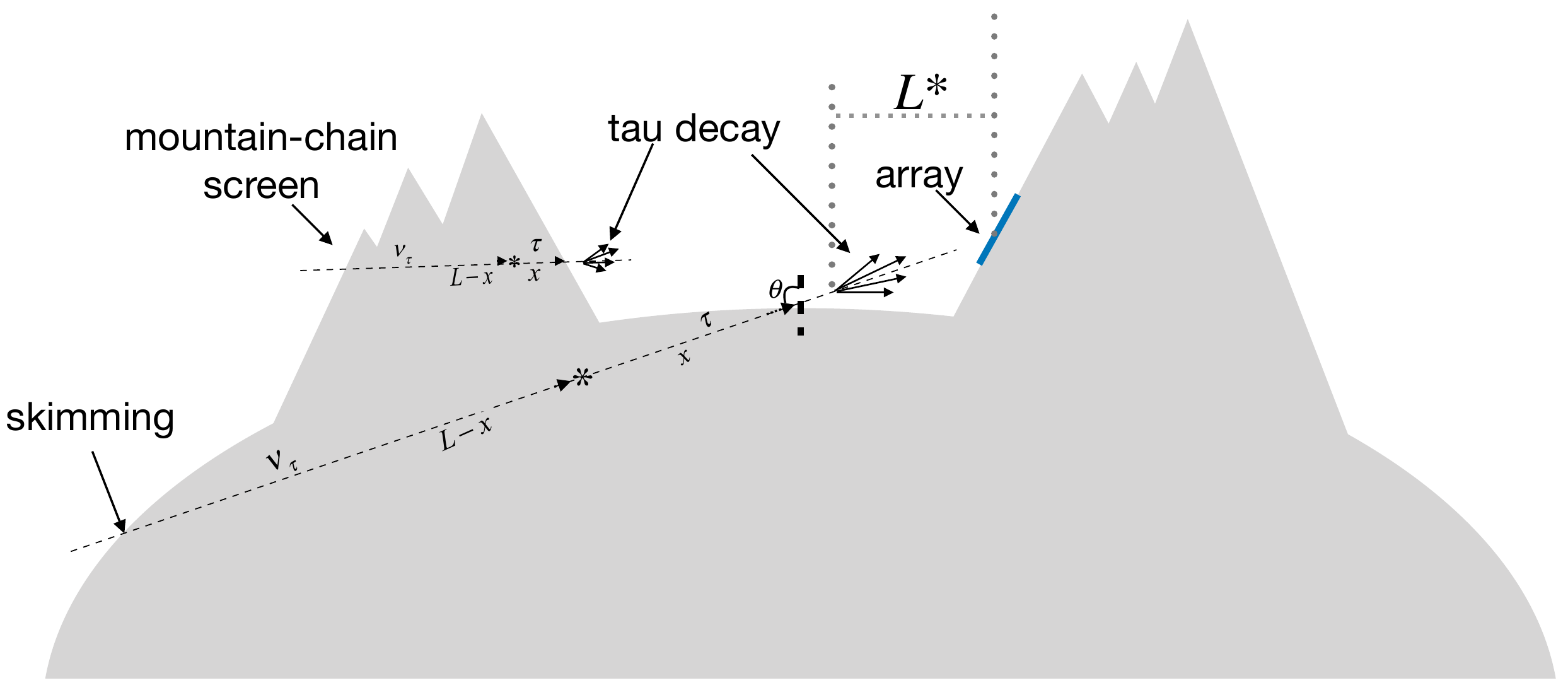}
  	\caption{In this layout (not in scale)  the array accepts $\tau$ shower produced by charged current interaction skimmed neutrinos by the Earth crust and  a  mountain in front of the  array.
  	$L-x$ is the interaction distance for the ${\nu}_{\tau}$, the interaction point shown by an asterisk symbol $``*"$,  $x$ is the traveling distance for the ${\tau}$. $L^{*}$ is the distance of $\tau$ decay point from the center of the array.}
	\label{fig:strategy}
\end{figure}

The detection of  high energetic ${\nu}_{\tau}$ which are coming nearly horizontal to the Earth
surface has a chance to create a $\tau$ via a charge-current
(CC) interaction or neutral-current (NC) interaction. It may survive and emerge from the surface to initiate a shower in the atmosphere.
In order to take advantage  of both valley and mountain to improve the detection probability,  each station in the array is placed on the mountain slope and directed to the horizon. The shower detection probability depends on not only $\tau$ initial energy but also traveling distance of the $\tau$ after the $\nu_{\tau}$ interaction inside the Earth crust. Therefore, we have considered a mountain in front with 60 km thickness which corresponds to maximum travel distance for $\tau$ with an energy of 1000 PeV  to escape from the rock before it decays as shown  in the sketch of the proposed experiment for this study (Figure~\ref{fig:strategy}). The incoming $\nu_{\tau}$ travels a distance $(L - x)$  along the Earth crust and 
interacts with the rock as depicted in the asterisk symbol, $``*"$ and produce $\tau$  by CC interaction. The $\tau$ travels a distance \textit{x} before emerging from the rock. $L^{*}$ is the distance between the point where $\tau$ initiates the shower and the center of the array. In this study, we have required that $\tau$ shower is produced in atmosphere.

To evaluate the expected $\tau$ events detected by the array we first  generated events 
using the TAUOLA code~\cite{JADACH1993361}
considering several decay modes of the $\tau$ in the energy range of 10-1000 PeV.
We have considered the decay modes ($\pi^{-}$, $\pi^{-}\pi^{0}$, $\pi^{-}\pi^{+}\pi^{-}$, $\pi^{-}\pi^{0}\pi^{0}$, $\pi^{-}\pi^{+}\pi^{-}\pi^{0}$) that covers 64 \% of the $\tau$ branching ratio to study the optimization and identification
performance of the expected $\nu_{\tau}$ induced showers by TAUWER array.

Consequently, CORSIKA (version 6.99) \cite{Heck:1998vt} has been modified and compiled to simulate very inclined 
showers at an observation plane which is considered to be $30\degree$. The produced air showers,
induced by the decay products of $\tau$, were initiated at 1500 m a.s.l. and developed up to the detector 
level (2250 m a.s.l.) for 4 different distances starting from the $\tau$ decay  point to the center of the array, $L^{*}$ (3 km, 5 km, 7 km and 10 km), as shown in Figure 2.  1000 showers have been simulated at  different energy for this study.
QGSJETII~\cite{Ostapchenko:2006aa} and GEISHA \cite{geisha} are selected  for high and low energy 
interaction models, respectively. ``CURVED EARTH" and ``SLANT" options are activated  to make
more realistic simulation for correctly include the atmospheric depth.
``COAST" option is selected for reading and converting CORSIKA binary files. 
Since the computing time of the showers increases with the energy, the ``thinning" strategy
has been also used in shower productions. The CORSIKA output file provides position ($x$, $y$ and $z$), arrival time and momentum ($P_x$, $P_y$, $P_z$, and $P_{tot}$) informations of each particle
crossing on the observation plane. Using these informations the trajectory is reconstructed for each 
particle and counted when it passes trough the scintillating tiles. In this proposed layout it is also important to identify the best distance of the plane of the array from the $\tau$ decay point to intercept the maximum shower evolution to make more efficient the shower detection and in the some time evaluate  the best valley wideness.

\begin{figure}[!htb] 
 	\centering 
    		\includegraphics[width=0.8\textwidth]{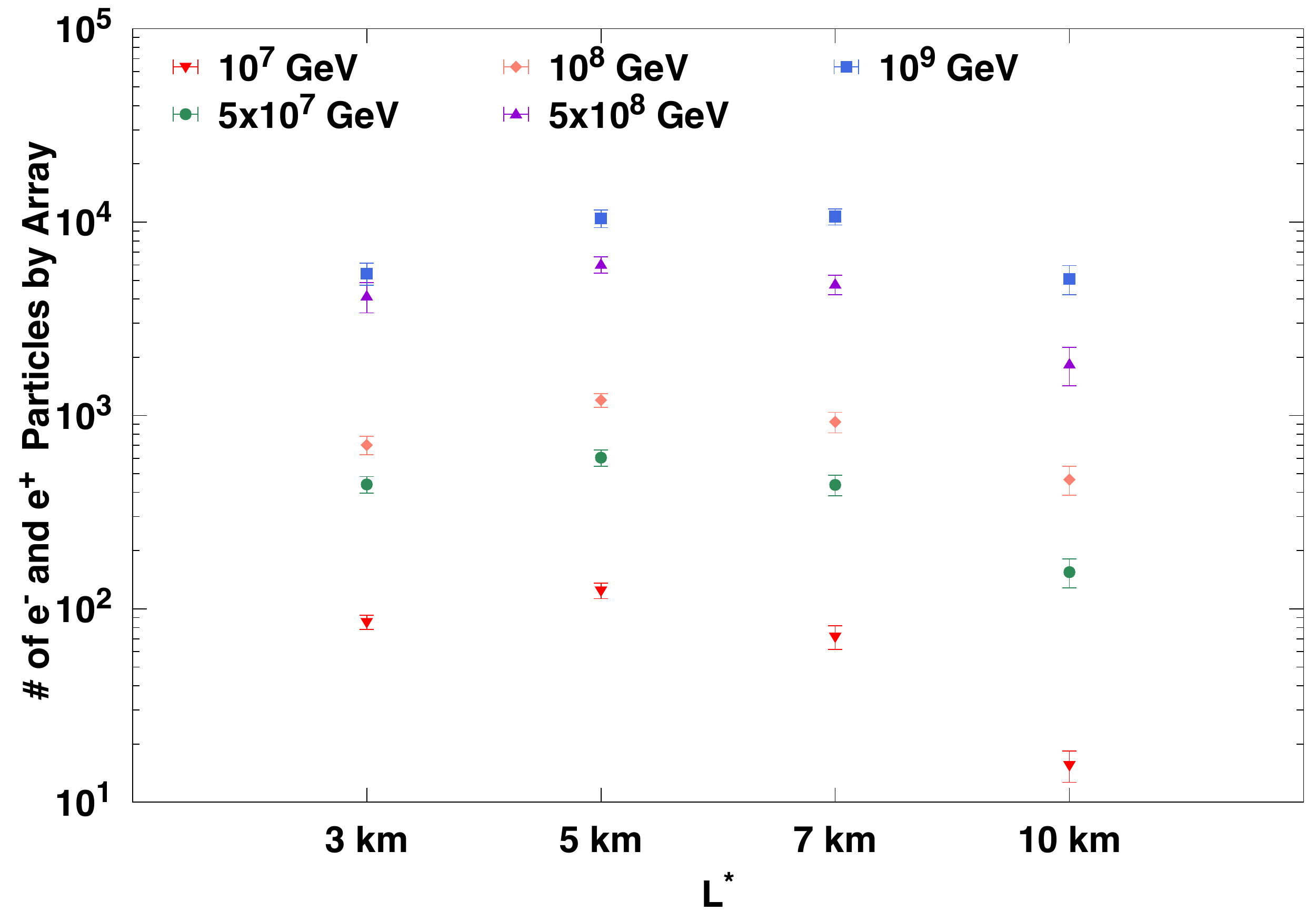}		
  	\caption{Averaged number of ${e^{\pm}}$ per shower (among 180 simulated showers) detected on the array as a function of $L^{*}$, distance of $\tau$ decay point to the center of the array, for different energies ($10^{7}$, $5x10^{7}$, $10^{8}$, $5x10^{8}$, $10^{9}$ GeV) at $\pi ^{-} \pi ^{0}$ decay mode.}
	\label{fig:motivation1_7km}
\end{figure}

 The determination of the optimum wideness for the valley is evaluated using the particle density measured on the array plane for different decay paths of  the $\tau$ in air. Figure~\ref{fig:motivation1_7km}  shows the averaged number of the ${e^{\pm}}$ per shower detected on the array plane for  $\pi ^{-} \pi ^{0} $ modes at different energies (10 - 1000 PeV) as a function of $L^{*}$, the distance of the $\tau$ decay point to the center of the array. The number of muons is about 5$\%$ of the number of electrons and almost flat as a function of the distance of $\tau$ decay point in the air.
It shows that the number of $e^{\pm}$ from the $\pi^{-} \pi^{0}$ decay mode and different $\tau$ energy  on the detector plane is maximum when the $\tau$ decay point is at a distance of 5-7 km from the center of the array. 
 A similar distribution is also obtained by  $\pi^{-}$,  $\pi^{-}\pi^{+}\pi^{-}$, $\pi^{-}\pi^{0}\pi^{0}$ and $\pi^{-}\pi^{+}\pi^{-}\pi^{0}$ decay modes.
 The decrement on the density of $e^{\pm}$ at 10 km is  due to the energy loss of $e^{\pm}$ in the atmosphere. 
 Since the density of the detected particles is higher at 5-7 km, 7 km decay lengths from emerging point to the array center is used for further analysis in this text. 
To obtain the expected number of neutrino $\tau$ detected by the array we have to evaluate the probability of escaping of the $\tau$ from the Earth crust, $P_{escp}$.
The probability of $\nu_{\tau}$ interacts  along the $L- x$ path  at $*$ point inside the rock is given as
 
 \begin{equation}
\label{eqn:Pint}
 P_{int} =  \int_{0}^{L-x}  e^{-l / L_{int}} \frac{dl}{L_{int}}
 \end{equation}
 where $L_{int}$ is the  neutrino  interaction length, $dl$ is the short distance the interaction happens. The neutrino cross section for charged current interaction within 10$\%$ as follows according to CTEQ6.6M~\cite{Jeong2017} and its energy dependence  varies as ${E_{\nu}}^{0.348}$ where $E_{\nu}$ is the $\nu_{\tau}$ energy. By this parameterization we have an interaction length of $\tau$ about 700 km for 10 PeV and 3000 km for 200 PeV. The neutrino NC cross section is 2.4 smaller than that of CC interaction which means the interaction length of neutrino is about two times greater than that of charged current channel. Since the NC interaction probability is no high before the CC interaction, the neutrino has the same energy as it enters the Earth shell when it interacts as CC. 
In this study we consider CC interactions of $\nu_{\tau}$ where the $\tau$ lepton is produced with 80\% of the $\nu_{\tau}$ energy~\cite{Gandhi:1998aa} as well as the NC interaction.

On the other hand, the probability the $\tau$ escapes from the Earth shell is given by the probability the neutrino interacts along its path  times the survival probabilities of $\tau$ lepton in the shell with an initial energy as a function of the distance it travels, given as

\begin{equation}
\label{eqn:Pescp} 
\resizebox{.9\hsize}{!}{$P_{escp} =P_{int}\times exp \Bigg [ -\displaystyle \int_{L-x}^{L}  ( c t_{0} \, ( exp [ -\beta_{0} / \beta_{1} +(\beta_{0}/\beta_{1} + \ln{(E_{i} / E_{0}}))] exp [-\beta_{1} \times \rho_{rock} \times l] E_{0})  / m_{\tau})^{-1}  dl \Bigg] $}
 \end{equation}
where $l$ is the distance the $\tau$ travelled, $c$ is the speed of light, $t_{0}$ is the mean life time of $\tau$ lepton, $\beta_{0}$ is the energy loss parameter from Bremsstrahlung, pair production and photonuclear interaction with $E_{0} = 10^{10}$ GeV, $E_{i}$ is the energy of $\tau$ lepton when it is generated, $\rho_{rock} = 2.65\times10^{15}$ (g/km$^3$)  is the density of the standard rock, and $m_{\tau}$ is the mass of $\tau$ lepton, $\beta_{1}$ is the photonuclear coefficient as function of neutrino energy~\cite{PhysRevD.81.114012, Jeong2017, Reno:2019qmk}. The maximum distance the $\tau$  lepton travels at the energy of 1000 PeV is about 60 km. That means the neutrino interaction should  happen in the Earth  shell of 60 km thickness to produce a tau lepton which has a probability to escape from the Earth crust.
 The survival probability ($P_{surv}$) of $\tau$, is correspond to the exponential factor in Eq.~\ref{eqn:Pescp}, decreases while $\tau$ lepton moves away from the interaction point in the Earth shell.

\subsection{Estimation on the number of detected ${\nu}_{\tau}$ events}
\label{sec:EventNumber}

As discussed in Section~\ref{sec:simulation}, we evaluated the $\nu_{\tau}$ interaction length and 
its  probability $P_{int}$ based on the study given in Ref.\cite{Gandhi:1998aa}, 
besides that the approaches are used from Ref.\cite{Dutta:2005aa} to calculate the $\tau$'s $P_{surv}$
and then the escape probability, $P_{escp}$ after the $\tau$ travels through the Earth and the mountain chain $60$ km thickness at PeV energies. Table~\ref{table:TableProbability} shows the $P_{escp}$ of the $\tau$ calculated for different neutrino initial energies and escaping directions from the Earth crust using the formula~\ref{eqn:Pescp}. The  $\tau$ crossing  the mountains are produced at $90\degree$ in $\theta$. No specific profile of the mountain has been studied.

\begin{table*}[!hbt]

\begin{center}
\caption{The escape probability of $\tau$  as function of  $\theta$ and $\nu_{\tau}$ energy. In the  first row is reported the probability considering a mountain screen 60 km thickness. These values where computed using a neutrino cross section for charged currents according to CTEQ6.6M.  \label{table:TableProbability}}
\begin{adjustbox}{max width=1\textwidth}
\rowcolors{1}{gray!15}{white!15}
\begin{tabular}{ccccccccc} \hline \hline
\multicolumn{1}{c}{$\theta$(\degree) / Energy (PeV)} & \multicolumn{1}{c}{10}&
\multicolumn{1}{c}{20} & \multicolumn{1}{c}{50} &
\multicolumn{1}{c}{100} & \multicolumn{1}{c}{200} &
\multicolumn{1}{c}{300} & \multicolumn{1}{c}{500} &
\multicolumn{1}{c}{1000}
\\
\hline
Screen & $6.850\times10^{-4}$ & $3.168\times10^{-3}$ & $9.430\times10^{-3}$ & $1.546\times10^{-2}$ & $2.220\times10^{-2}$ & $2.651\times10^{-2}$ & $3.245\times10^{-2}$ & $4.164\times10^{-2}$ \\
$91$ & $3.114\times10^{-6}$ & $2.038\times10^{-5}$ & $9.056\times10^{-5}$ & $1.900\times10^{-4}$ & $3.276\times10^{-4}$ & $4.203\times10^{-4}$ & $5.381\times10^{-4}$ & $6.669\times10^{-4}$ \\
$92$ & $3.799\times10^{-6}$ & $2.149\times10^{-5}$ & $7.418\times10^{-5}$ & $1.212\times10^{-4}$ & $1.525\times10^{-4}$ & $1.570\times10^{-4}$ & $1.456\times10^{-4}$ & $1.061\times10^{-4}$ \\
$93$ & $3.383\times10^{-6}$ & $1.659\times10^{-5}$ & $4.472\times10^{-5}$ & $5.729\times10^{-5}$ & $5.323\times10^{-5}$ & $4.434\times10^{-5}$ & $3.023\times10^{-5}$ & $1.330\times10^{-5}$ \\
$94$ & $2.670\times10^{-6}$ & $1.138\times10^{-5}$ & $2.408\times10^{-5}$ & $2.435\times10^{-5}$ & $1.686\times10^{-5}$ & $1.145\times10^{-5}$ & $5.806\times10^{-6}$ & $1.571\times10^{-6}$ \\
$95$ & $1.978\times10^{-6}$ & $7.340\times10^{-6}$ & $1.226\times10^{-5}$ & $9.851\times10^{-6}$ & $5.124\times10^{-6}$ & $2.855\times10^{-6}$ & $1.084\times10^{-6}$ & $1.825\times10^{-7}$ \\
$96$ & $1.411\times10^{-6}$ & $4.571\times10^{-6}$ & $6.056\times10^{-6}$ & $3.883\times10^{-6}$ & $1.527\times10^{-6}$ & $7.009\times10^{-7}$ & $2.006\times10^{-7}$ & $2.114\times10^{-8}$ \\
$97$ & $9.818\times10^{-7}$ & $2.784\times10^{-6}$ & $2.938\times10^{-6}$ & $1.510\times10^{-6}$ & $4.515\times10^{-7}$ & $1.712\times10^{-7}$ & $3.704\times10^{-8}$ & $2.453\times10^{-9}$ \\
$98$ & $6.709\times10^{-7}$ & $1.667\times10^{-6}$ & $1.405\times10^{-6}$ & $5.810\times10^{-7}$ & $1.324\times10^{-7}$ & $4.150\times10^{-8}$ & $6.798\times10^{-9}$ & $2.830\times10^{-10}$ \\
\hline \hline
\end{tabular}
\end{adjustbox}
\end{center}
\end{table*}

By using the present cosmogenic neutrino flux upper limit~\cite{IceCube2018, 2012Icecube, :2016aa}, $\Phi(E)$,  we evaluate the expected number of $\tau$  events in the TAUWER array according to the following formula

\begin{equation}
\label{Neq1}
 \resizebox{.9\hsize}{!}{$N_{\tau} = \int_{E_{min}}^{E_{max}} \Phi(E)P_{escp}(E) (\varepsilon_{acc}(E,\Theta, \Phi, L^*)\times\beta_{\tau} )dE \Delta t \Delta A \Delta \Omega$}
\end{equation}
where $\beta_{\tau}$ is $\tau$ branching ratios for $\pi^-$, $\pi^-\pi^0$, $\pi^-\pi^+\pi^-$, $\pi^-\pi^0\pi^0$ and 
$\pi^-\pi^+\pi^-\pi^0$ decay modes, $\varepsilon_{acc}$ is the efficiency of detector array acceptance. $\Delta t$ is the time interval of one year, $\Delta A=0.6$ $km^2$ area  and $\Delta \Omega$ is the solid angle that watched the showers emerging below the horizon within $8 \degree$ in $\theta$  and the showers emerging from the chain of mountain in front of the array of $2\degree$ in $\theta$ .
To evaluate the number of detected events, we take into account two scenarios: Earth Skimming and the screen strategies
produced by a mountain chain in front of the array.

The expected number of events by the TAUWER array was estimated using the latest neutrino flux upper limit reported by IceCube experiment~\cite{IceCube2018,2012Icecube,:2016aa}. The results of different scenarios, only skimming strategy and screen strategy are summarized in Table~\ref{table:TableExpected2}. 
The expected number of events as function of two  different neutrino flux  upper limits  are $0.18$ and $0.43$ events in $km^{2}$ per year for skimming and skimming-screen strategy, respectively. 
If the integrated time is 3 years \cite{:2016aa} and the surface of array is 10 $km{^2}$  the number of expected events  for TAUWER array are about $13$ events.
The presence of `screen' increases the detected number of events about a factor three. That is explained by higher probability of interaction of the $\nu_{\tau}$ that travels a longer path when it is skimmed by the Earth crust.

\begin{table*}[!hbt]

\begin{center}
\caption{The expected number of events in $km^{2}$ per year for TAUWER array using analytic method and  DANTON simulation code corresponding to the different upper limit fluxes.  With DANTON simulation we computed only skimming events.\label{table:TableExpected2}}
\begin{adjustbox}{max width=1\textwidth}
 \begin{tabular}{ccc >{\columncolor{white!15}} c >{\columncolor{white!15}} c} \hline \hline
 \multicolumn{1}{c}{ Flux } &\multicolumn{1}{c}{Scenario} & \multicolumn{1}{c}{Energy range} & \multicolumn{1}{c}{$N_{event}$} & \multicolumn{1}{c}{$N_{event}$}   \\ 
  \multicolumn{1}{c}{ ($GeV cm^{-2}s^{-1}sr^{-1}$)  } &\multicolumn{1}{c}{ } & \multicolumn{1}{c}{ (PeV)} & \multicolumn{1}{c}{CTEQ6.6M } & \multicolumn{1}{c}{DantonSim}   \\ \hline

 &Skimming & $[10 - 1000]$  & 0.0437 & 0.0885   \\
  $2\times10^{-8}$ (all flavor) \cite{IceCube2018} &Screen & $[10 - 1000]$  & 0.1273 & $ $ \\
 &Skimming + Screen & $[10 - 1000]$ & 0.1710 & $ $   \\ \hline
 &Skimming & $[10 - 1000]$ & 0.1116 & 0.2257 \\
  $5.1\times10^{-8}$ ($\nu_{\tau}$ flavor) \cite{:2016aa} &Screen & $[10 - 1000]$ & 0.3245 & $ $  \\
 &Skimming + Screen & $[10 - 1000]$   & 0.4361 & $ $  \\
\hline \hline
\end{tabular}
\end{adjustbox}
\end{center}
\end{table*}
 The  $P_{escp}$ computed using the analytic method has been compared to that evaluated using a specific code DANTON  that simulates the tau-neutrino interaction in the Earth crust and the $\tau$ decay giving the $\tau$ decay point in the air~\cite{Niess:2018opy}. Using this code we obtained the total $P_{escp}$ of $\tau$ for energy greater than 200 PeV in average a factor 2.5 higher than the $P_{escp}$ evaluated with the analytic method. This discrepancy is due mainly to the different PDF used in the neutrino cross section parametrization.  To evaluate the expected $\tau$ lepton we can use the geometrical acceptance computed in our study because the  tau decay point coordinates in DANTON have  an average distance from the array of 8 km and a momentum distribution similar to which was used  in the analytic method. By using the upper limit fluxes~\cite{IceCube2018,:2016aa}, we obtain 0.21 and 0.55 events per km$^2$ per year with the DANTON simulation code.

\subsection{Arrival $\tau$ shower  Direction}
\label{sec:ArrDirection}

\begin{figure}[h]
 \centering 
  \includegraphics[width=1\textwidth]{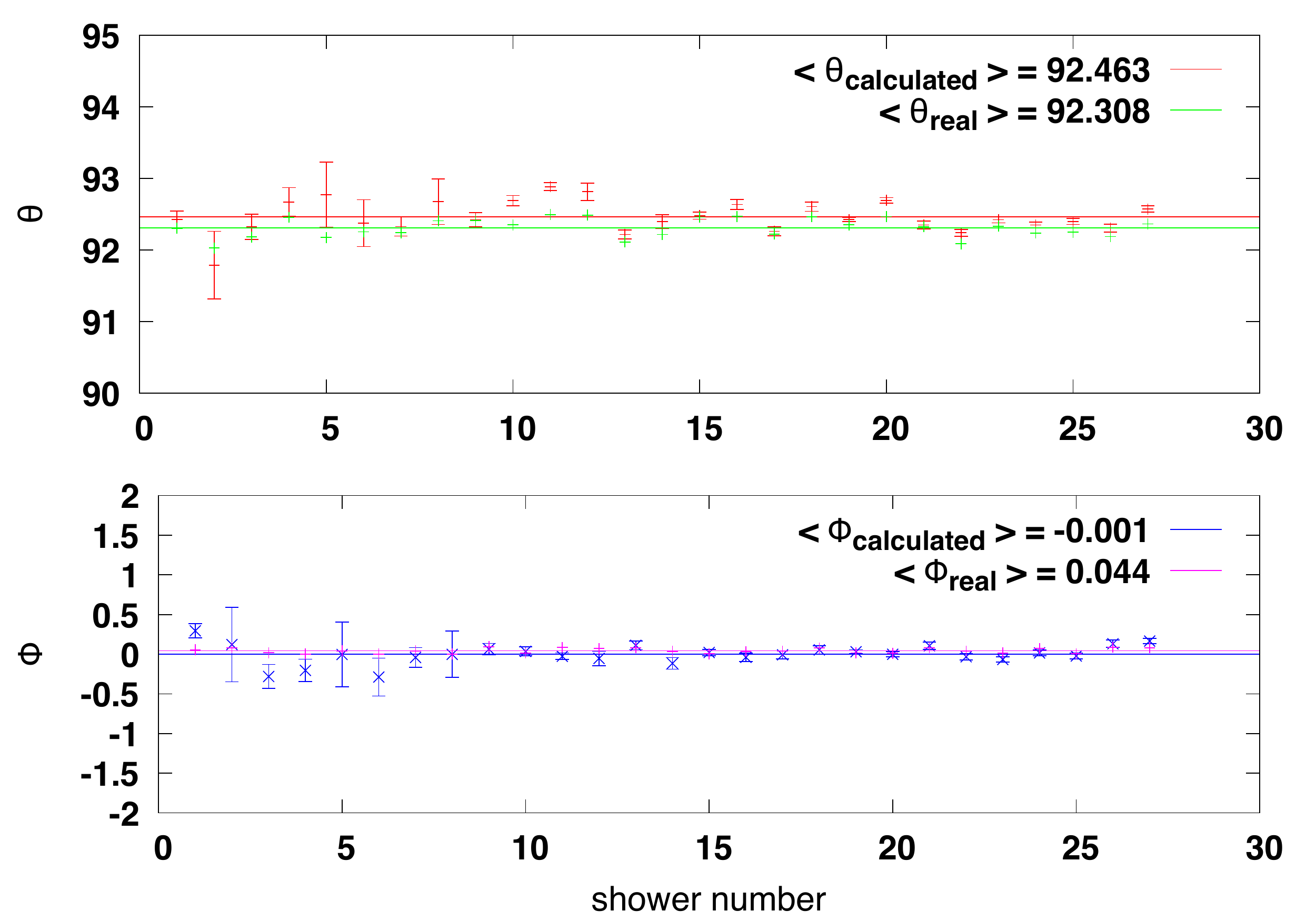}
  \caption{ Zenith (top) and Azimuthal (bottom) angles evaluated from active station for different showers 
  where decay length from emerging point is 7 km and primary energy ranges from $10$ PeV to $1000$ PeV. Active stations are 
  selected by requiring at least two hits on the tower.} 
 \label{fig:directionAngle}
\end{figure}

The direction of the $\tau$ air shower is calculated with a common method used 
in literature \cite{Hedayati} based on the minimization of the direction vectors of each 
track  by using Minuit \cite{Minuit}.
The momentum  of each particle are used to obtain  the direction vectors $(n_{x}$, $n_{y}$, $n_{z})$ of the shower in Eq.~\ref{eq:chi}.

\begin{eqnarray}
    \label{eq:chi}
    \left.\begin{aligned}
     \chi^2 = \sum_{i}^{N} w_{i}\Big(& n_{x}(x_{i}-x_{core}) + n_{y}(y_{i}-y_{core})  \\ 
     & + n_{z}(z_{i}-z_{core}) - c(t_{i}-t_{core}) \Big)^2 
 \end{aligned}\right.
\end{eqnarray}
where $t_{i} $ is the arrival time of $e^{\pm}$ , $x$, $y$ and $z$ its coordinates on the array plane,
$x_{core},\, y_{core},\, z_{core}$ the coordinates of the core of the shower evaluated with maximum weighted density  
of the particles on the inclined plane. $t_{core}$ is estimated from the closest station taken into account of the distance from the core. 
The active stations  are selected by requiring at least two particles on the \textit{tower} and used 
in the evaluation of the shower direction for different showers where the primary 
in an energy range of $10$ PeV - $1000$ PeV and the decay point 7 km distant from the array plane.

Figure~\ref{fig:directionAngle} shows both zenith and azimuthal angles evaluated for sample of $27$ showers in the energy range between 10 and 1000 PeV.
The arrival directions $\theta$ and $\phi$ are estimated with an error of $0.16\degree$ and $0.045\degree$,
respectively. 

\subsection{$\tau$ shower energy reconstruction}
\label{sec:EnergyReconstruct}

\begin{figure}[h] 
 \centering 
\subfigure[]{%
 \includegraphics[width=0.5\textwidth, height=0.27\textheight]{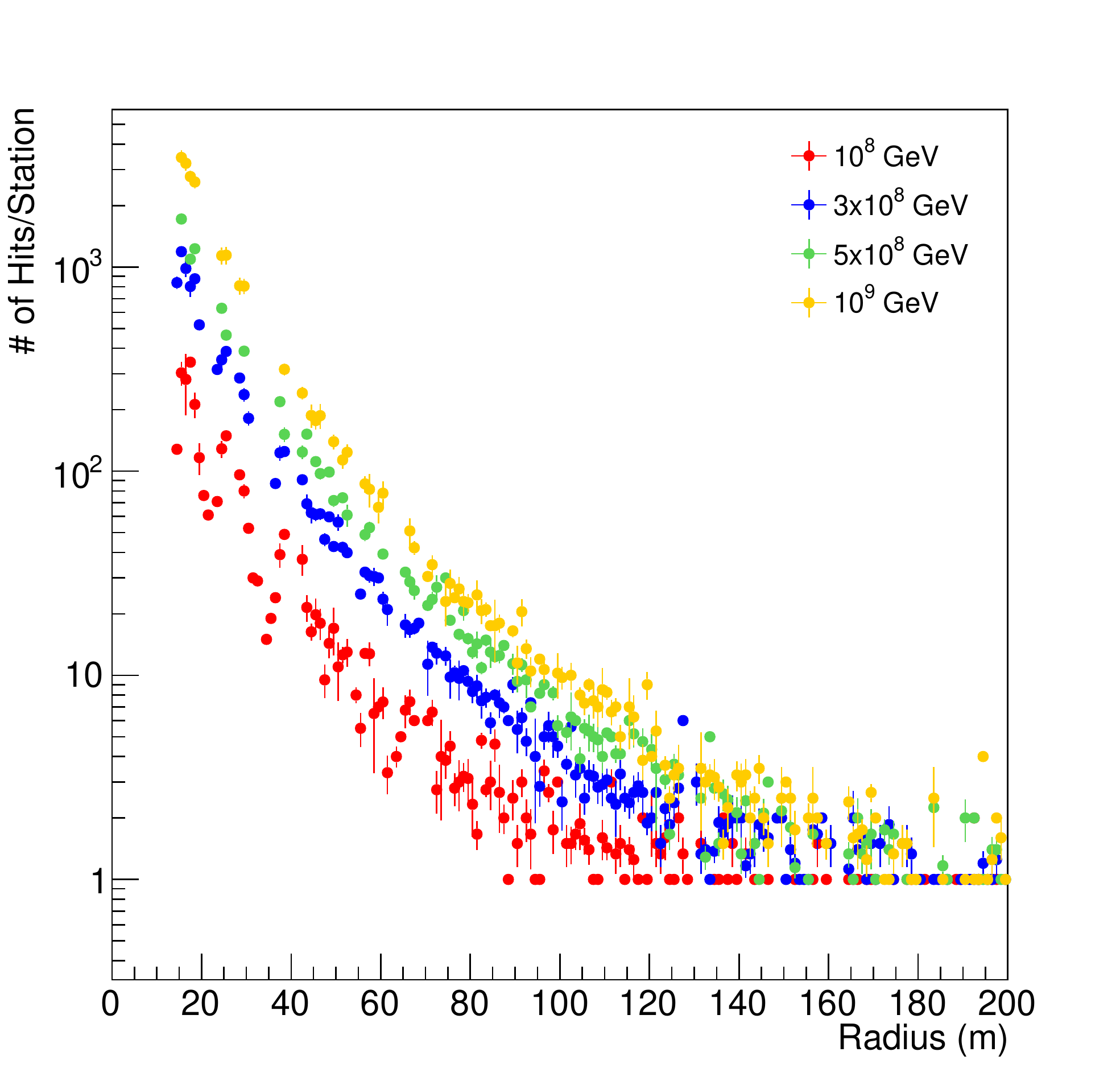}
\label{fig:hit_rad1}
  }
	\subfigure[]{%
    		\includegraphics[width=0.47\textwidth, height=0.25\textheight]{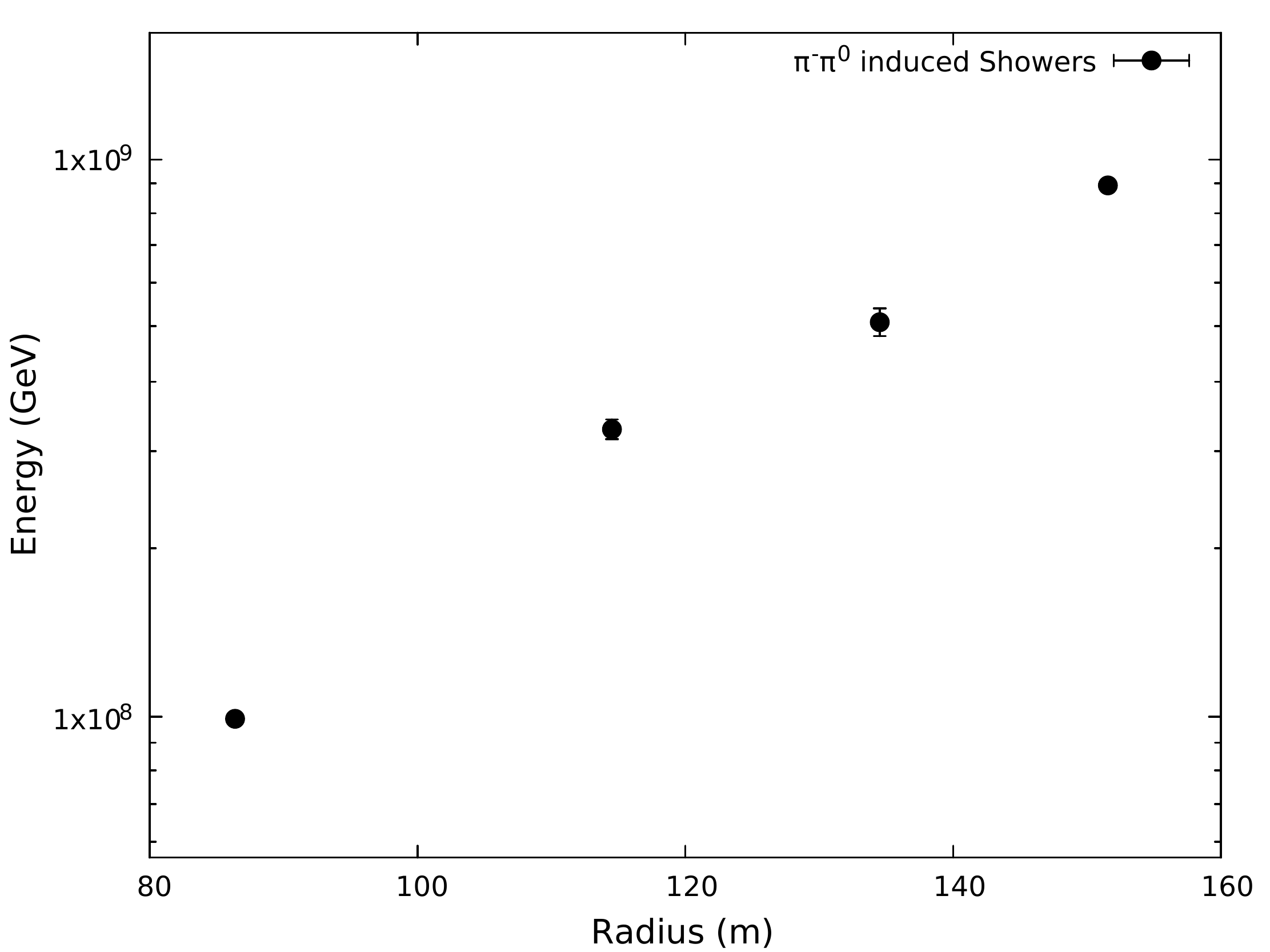}
		\label{fig:e_radius1}
    	}		
\caption{a) Number of hits per station versus distance from the shower core for different $\tau$ energies,
b) the energy of $\tau$ showers as function of the shower radius.}
\label{fig:radius}
\end{figure}

The center of the shower is estimated from the most triggered stations and 
hits per station as function of distance from the shower core is given in Figure~\ref{fig:hit_rad1}.
This figure shows the number of hits per station as a function of distance for different energy intervals between 10 and 1000 PeV where at least one hit on a single tile is required. 
The core of the shower is evaluated from the stations having at least 5 hits. While the radius of the shower is obtained by requiring 4 subsequent stations having hits ranging from 1 to 4 and
register the first one as the radius (edge) of the shower. The Figure~\ref{fig:e_radius1}
shows the energy versus its corresponding radius. It is clearly seen that there is an exponential behaviour between energy and radius in the range of the interested energy region.

\subsection{Trigger Decision}
\label{sec:TriggEff}

The average timing spectrum for $e^{\pm}$ between 10 PeV and 1000 PeV for all hadronic channels of $\tau$ shower that hit any scintillator in the 640 stations has been studied.
The timing spread of the stations is computed by using the arrival time of all hits in the station and identifying the earliest one which is selected as time reference to find a delay time in a $(4\times4)$ subarray around the center of the shower. The maximum time difference between the first and last hit produced by the $\tau$ shower on the array plane must be in the time interval of $0.3\,\mu s$. From these simulation studies the signature of $\tau$ shower is given by a cluster of station with highest density of $e^{\pm}$ and the maximum number of hits in a single station for $\tau$ shower between 10 PeV and 1000 PeV ranges from $450-3500$, in the time interval of $10-20$ ns. Compared this requirement to the vertical EAS events collected in a test made at Karlsruhe Institute 
of Technology (KIT) \cite{IORI2012285},  the atmospheric background can be easily suppressed if
we require $250$  hits, defined as trigger T1. The second level trigger T2 can be defined as requiring at least $450$ hits on each station of a sub-array $(2\times2)$ in a time
window of $88$ ns. This criteria (/trigger) leads to a selection of a shower with an energy of $10^8\, GeV$ or greater.

\section{Conclusions}
\label{sec:Conc}

We have simulated $\tau$ decay modes ($\pi^{-}$, $\pi^{-}\pi^{0}$, $\pi^{-}\pi^{+}\pi^{-}$, $\pi^{-}\pi^{0}\pi^{0}$, 
$\pi^{-}\pi^{+}\pi^{-}\pi^{0}$) that covers 64 \% of the $\tau$ branching ratio
in the energy range of $10-1000$ PeV to determine the performance of the TAUWER array.
It shows that the energy of the showers between $100$ and $1000$ PeV can be reconstructed in terms of the radius 
as discussed in Section~\ref{sec:EnergyReconstruct}.
The arrival direction is also determined with an accuracy of $0.16^o$ in the same energy range. The atmospheric background can be easily rejected by requiring T1 trigger while a second level trigger, T2 selects showers 
with $10^8$ GeV or greater energies. We have estimated the $\nu_{\tau}$ conversion to $\tau$ and $\tau$ propagation through the rock including both skimming and screen strategies with the TAUWER detector array,
which is ranging between  $5$-$13$ tau lepton events in the energy range of $10-1000$ PeV for $3$ years of integrated time period and an array surface of 10 $km^{2}$ using the analytical method. 
We obtained $6$-$16$ tau lepton skimmed events by using DANTON simulation for the same time period, array surface and in same  energy range. The number of events skimmed by the Earth crust using this simulation code is higher of a factor two due to the neutrino cross section. The effect of the screen in front of the array increase the expected number of events of a factor three. 

\section{Acknowledgement}

We acknowledge the financial support by the Scientific and Technological Research Council of Turkey (TUBITAK)
(grant number: 114F138).

\bibliographystyle{ieeetr}
\bibliography{mybibfile}

\end{document}